\documentclass[12pt]{article}
\usepackage{amssymb,psfrag,graphicx}
\title{Marginally outer trapped surfaces in higher dimensions \thanks{Preprint UWThPh-2013-7}}
\author{Tim-Torben Paetz \thanks{tim-torben.paetz@univie.ac.at}~ and Walter
Simon \thanks{walter.simon@univie.ac.at}}
 \date{\normalsize Gravitationsphysik, Fakult\"at f\"ur Physik, Universit\"at
 Wien\\ Boltzmanngasse 5, A-1090 Wien, Austria}
\begin{document}
\maketitle
\begin{abstract}
We review the basic setup of Kaluza-Klein theory, namely a 5-dimensional vacuum with a
cyclic isometry, which corresponds to Einstein-Maxwell-dilaton theory in
4-dimensional spacetime. We first recall the behaviour of Killing horizons 
under bundle lift and projection. We then show that the
property of compact surfaces of being (stably) marginally trapped is
preserved under lift and projection provided the appropriate ("Pauli-") 
conformal scaling is used for the spacetime metric. We also discuss and
compare recently proven area inequalities for stable axially symmetric 
2-dimensional and 3-dimensional marginally outer trapped surfaces.    
\end{abstract}

\section{Introduction}
The basic setup of Kaluza-Klein (KK) theory consists of a 4+n dimensional manifold with n Killing vectors, 
which can be regarded as 
a principal fibre bundle over a 4d base (see e.g. \cite{OW,YC1}). In 5 dimensions in particular,
stationary black holes have been studied extensively and revealed much richer 
structures than in 4d (cf e.g. \cite{ER}--\cite{FL}). There is, however, a more subtle motivation 
for studying black holes in (4+n) dimensions: Suitable adaptions of Hawking's rigidity 
theorem assert the existence of n+1 spacelike isometries 
on generic (non-static) stationary event horizons \cite{HI1}, and under 
suitable technical assumptions, such Killing fields can be 
extended off the horizon. From this point of view, the higher dimensional
black hole is the "fundamental" object which by itself is capable of 
"generating 4d spacetime" with matter fields, via n isometries which 
serve as KK-fibres \cite{WS}. In addition, the horizon "generates" axial symmetry as usual.

In more realistic KK theories the n fibre-isometries are replaced  
by periodicity conditions on all fields involved (cf e.g. \cite{OW}). This raises the question
whether such structures could arise "automatically" as well, namely from a non-stationary but suitably structured 
horizon, as a consequence of a hypothetical generalized rigidity theorem. While
a concrete implementation of this speculation would lead beyond the scope of the 
present work, the basic idea does give a motivation for  studying general gravitational collapse in higher 
dimensions, as initiated below.   

The key concepts in the study of gravitational collapse are  (outer) trapped surfaces
and marginally outer trapped surfaces (MOTS). The latter are defined as compact 
codimension 2-surfaces on which the orthogonally outgoing family of null geodesics 
has vanishing expansion (cf e.g. \cite{JLJ1}). In applications, "stable" and "strictly stable" MOTS play a 
distinguished role and have been studied extensively \cite{JLJ1}--\cite{IC}. Here stability is a mild restriction 
which roughly speaking requires the existence of a codimension 1-neighbourhood 
in a selected normal direction, which can be foliated by trapped surfaces inside and untrapped surfaces outside. 
On the other hand, it has been shown that the (global) boundary of the trapped region in a
spacelike hypersurface is formed by a smooth, stable MOTS \cite{AM,ME,AME}. Originally, the stability condition
came up in connection with the topology theorems for MOTS \cite{HE,RN,GG}. More recently, however,
the crucial role of stability in the time evolution of MOTS in a foliated spacetime was clarified, 
as stability implies smooth evolution while instability signals "jumps" \cite{AMMS}.
Moreover, stability is significant for the singularity theorems, as the original requirement of the existence of a 
trapped surface (both null expansions converging) can in essence be replaced by
the requirement of existence of a stable MOTS \cite{IC}. 
All these results apply in principle in higher dimensions, (at least till 7), and motivate our 
interest in the stability condition in particular. 

A further motivation comes from the area inequalities for stable MOTS \cite{JRD,SD,JLJ2}.  Such inequalities have been found in 
particular for axially symmetric, stable 2d MOTS in 4d Einstein-Maxwell by Gabach Clement, Jaramillo and Reiris \cite{GCJ, GJR} 
and in 4d Einstein-Maxwell dilaton (EMD) theory by Yazadjiev \cite{SY}. Other inequalities have been obtained for stable 3d MOTS with 
various topologies and corresponding symmetries in 5d spacetimes by Hollands \cite{SH}. Since a 5d vacuum with a KK 
symmetry is equivalent to EMD theory in 4d, the results of these papers can be compared provided the concepts of stable
2d MOTS and 3d MOTS can be matched. These are the main goals of the present paper.      

As we want to illustrate the basic concepts rather than investigating a physically realistic theory, we restrict 
ourselves to 5d vacuum spacetimes and their dimensional reduction along an isometry. We first (in Sect. 2) review 
Geroch's projection formalism \cite{RG} and continue in Sect. 3.1 with recalling the known behaviour of
Killing horizons and their generators under such a reduction \cite{WS,HI1}.  In Sect. 3.2 we show that 3d MOTS do project down to 2d ones
provided an appropriate conformal scaling (which may be ascribed to Pauli \cite{WP}) is used for the spacetime
metric. The scaling is irrelevant only in a degenerate case in which the norm of the KK Killing field is constant 
along the outgoing null direction.  
Fortunately, the Pauli-scaling is the same which is compulsory for matching the variational formulations in different 
dimensions anyway \cite{OW,YC2}. Next (in Sect. 3.3) we relate the stability definitions for 3d and 2d MOTS, which
is straightforward in principle but involves some subtleties.  
In  Sect. 4 we finally analyze and relate the area 
inequalities. This requires a careful discussion of the parameters and of the non-trivial topologies which arise 
in the presence of a magnetic monopole.

\section{Kaluza-Klein a la Geroch} 
We consider a smooth 5-dimensional Ricci-flat manifold  
$({}^5{\cal M}, {}^5g_{AB})$ of signature (- + + + +), 
where capital latin indices run from $0$ to $4$. 
We assume that $({}^5{\cal M}, {}^5g_{AB})$ is a principal fibre bundle  
with spacelike $U(1)$ isometry ${\cal C}$ as fibre and smooth base manifold ${\cal M}$
(cf. \cite{YC1}). This entails that ${\cal C}$ has no fixed points; the corresponding Killing 
field on ${}^5{\cal M}$ is denoted by $c^A$, i.e. the corresponding Lie
derivative  satisfies ${\cal L}_c {}~^5g_{AB} = 0$.
 The squared norm of $c^A$ is denoted by $V^2 = c^Ac_A > 0 $. 
 Subsets  of ${}^5{\cal M}$ are distinguished from subsets of ${\cal M}$  
by the superscript ${}^5$ (irrespective of their dimension). Before introducing a coordinate system in Equ. (\ref{met}), tensorial objects carry 
(Penrose's) "abstract indices " to indicate their nature; in particular, tensors on ${}^5{\cal M}$ carry  capital
 indices while those on ${\cal M}$ carry lower case indices ranging from  $0$ to $3$. 
A set ${\cal E} \subset {}^5{\cal M}$ is called {\bf invariant} if ${\cal C}({\cal E}) = {\cal E}$.  

Below we introduce a terminology which is some sort of amalgamation between standard fibre 
bundle language and Geroch's projection formalism \cite{RG}; 
However, for sets the terminology ``lift'' and ``projection'' 
used below is not standard, while for vectors we drop the word ``horizontal''
from the standard term ``horizontal lift''; see the remark at the end of 
this section.

\begin{figure}[h!]
\begin{psfrags}
\psfrag{M}{\Huge ${\cal M} $}
\psfrag{5M}{\Huge ${}^5{\cal M}$}
\psfrag{x}{\Huge$x$}
\psfrag{y}{\Huge $y = {\cal C}^{\uparrow}(x)$}
\psfrag{f}{\Huge $f $}
\psfrag{F}{\Huge $F = {\cal C}^{\uparrow}_*(f)$}
\psfrag{R}{\Huge $\mathbb{R}$}
\psfrag{Fy}{\Huge $F(y)= f \circ {\cal C}^{\downarrow}(y)$}
\includegraphics[angle=0,totalheight=4cm]{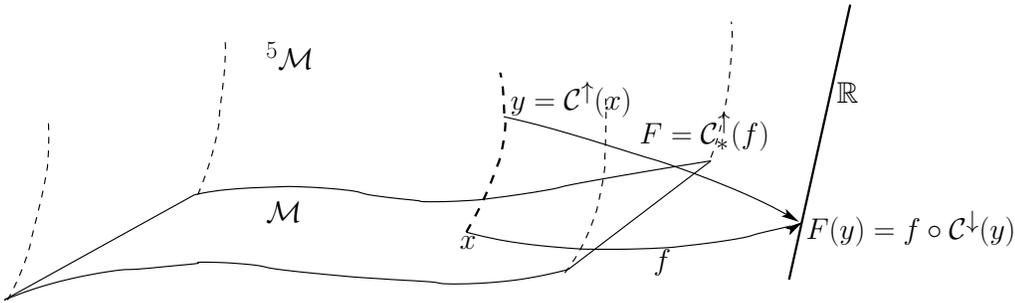}
\end{psfrags}
\caption{Lifts and projections of functions} 
\label{lpf}
\end{figure}

\begin{description}
\item[The lift ${\cal C}^{\uparrow}$ of a set ${\cal D}$:]~ \\
${\cal C}^{\uparrow}: {\cal M} \rightarrow {}^5{\cal M}$
assigns to a subset ${\cal D} \subset {\cal M}$ the set of points
${\cal E} \subset {}^5{\cal M}$ on the orbits through ${\cal D}$. 
\item[The projection ${\cal C}^{\downarrow}$ of an invariant set ${\cal E}$:]~
\\
 ${\cal C}^{\downarrow}: {}^5{\cal M} \rightarrow {\cal M}$ assigns to an invariant subset 
${\cal E} \subset {}^5{\cal M}$ the set of orbits ${\cal E}/{\cal C}$. 
\end{description}
These maps induce maps between {\bf invariant tensor fields}
$w^{A...M}_{~~~~~N...Z}$ on ${}^5{\cal M}$, defined by
\begin{equation}
\label{inv}
{\cal L}_c w^{A...M}_{~~~~~N...Z} = 0, ~~~~w^{A...M}_{~~~~~N...Z}c_A = 0
~~...~~
w^{A...M}_{~~~~~N...Z}c^Z = 0  
\end{equation}
 and tensor fields on ${\cal M}$. The maps displayed in
Fig.  \ref{lpf} and defined below are of course nothing  but "pullbacks" and "pushforwards".  
\begin{description}
\item[Lift ${\cal C}^{\uparrow}_*$ of a function $f: {\cal M} \rightarrow \mathbb{R}; f \in {\cal
F}({\cal M})$] (the module of functions) {\bf to an invariant function}~
$F: {}^5{\cal M} \rightarrow \mathbb{R}; F \in {\cal F}_-({}^5{\cal M})$  (the module of
invariant functions):~\\
$ {\cal C}^{\uparrow}_* :  {\cal F}({\cal M}) \rightarrow {\cal F}_-({}^5{\cal M})$
is defined as $F = {\cal C}^{\uparrow}_*(f) = f \circ {\cal
C}^{\downarrow}$.
 \item[Projection of an invariant function $F \in {\cal F}_-({}^5{\cal M})$:] ~ \\ 
$ {\cal C}^{\downarrow}_*: {\cal F}_-({}^5{\cal M}) \rightarrow 
{\cal F}({\cal M})$ is defined as $f = {\cal C}^{\downarrow}_*(F) = F \circ {\cal
C}^{\uparrow}$.    
\end{description}
The maps introduced above can now be extended straightforwardly to vector
and tensor fields by considering gradients of functions etc.. For definitions cf Appendix A of 
\cite{RG} - here we just fix notations:
\begin{description}
\item[Lift ${\cal C}^{\uparrow}_*$ of a vector field $w^i \in  T({\cal M})$
to an invariant vector field $z^A \in T_-({}^5{\cal M})$ :]~\\
(the  horizontal subspace of $T ({}^5{\cal M})).$\\
$ {\cal C}^{\uparrow}_* :  T({\cal M}) \rightarrow T_-({}^5{\cal M})$
 denoted by  ${\cal C}^{\uparrow}_*(w^i) = z^A$. 
 \item[Projection of an invariant vector field  $z^A$:]~\\ 
$ {\cal C}^{\downarrow}_*: T_-({}^5{\cal M}) \rightarrow T({\cal M})$ denoted by
 ${\cal C}^{\downarrow}_* (z^A) = w^i$.  
\end{description}
 
From the construction it is clear that ${\cal C}^{\uparrow}$ and ${\cal C}^{\downarrow}$ are not isomorphisms, 
while ${\cal C}^{\uparrow}_*$ and ${\cal C}^{\downarrow}_*$ are \cite{RG},
 which is indicated by the asterisks. 
 In what follows ${\cal C}^{\downarrow}_*$ will only be applied to invariant objects.
 ${\cal C}^{\uparrow}_*$ and ${\cal C}^{\downarrow}_*$ are homomorphisms in
 the sense that they cut through tensor products. 
In Lemma 1 below, the action of these mappings on covariant and contravariant vector fields will be displayed 
explicitly in the coordinate system (\ref{met}) adapted to the isometry.

The projection maps introduced above should not be confused with a particular tensor field called "the projector"
 defined by
\begin{equation}
\label{pr}
P^A_{~C} = \delta^A_{~C} - V^{-2} c^A c_C, 
\end{equation}
which can in fact be applied to {\it any} tensor field in $\otimes T({}^5{\cal M})$.
In particular, applied to ${}^5g_{AB}$ it gives the invariant metric
\begin{equation}
\label{g}
\widetilde g_{CD} = P^A_{~C}  P^B_{~D} ~{}^5 g_{AB},   
\end{equation}  
which effectively lives on ${\cal M}$; in our notation this reads 
${\cal C}^{\downarrow}(\widetilde g_{AB}) =  \widetilde g_{ij}$.
We also recall the volume forms $\widetilde \mu = * 1$ on ${\cal M}$ and 
${}^5\mu =  {}^{5}* 1$ on ${}^5{\cal M}$ where $*$ and ${}^5*$ are the Hodge
duals, and the antisymmetric symbols  $\widetilde \epsilon_{ijkl}$ with 
$\epsilon_{0123} = \sqrt{\mbox{det}~ \widetilde g}$ and $\epsilon_{ABCDE}$ with 
$\epsilon_{01234} =  \sqrt{\mbox{det}~ {}^5g}$. 
These objects are related via ${\cal C}^{\uparrow}_* (\widetilde \epsilon_{ijkl}) = V^{-1} \epsilon_{ABCDE}c^E$. 

The Levi-Civita conection w.r.t. ${}^5g_{AB}$ is denoted by ${}^5 \nabla_A$. 
For an invariant vector field $w^A$ another covariant derivative
$\widetilde \nabla_A$ can be defined by \cite{RG}
\begin{equation}
\label{cd}
\widetilde \nabla_C w_D = P^A_{~C}  P^B_{~D} ~{}^5\nabla_A w_B.   
\end{equation}
The derivative $\widetilde \nabla_A$ is metric with respect to (\ref{g}), torsion
free and $\widetilde \nabla_C w_D$ is invariant in the sense of (\ref{inv}). 
Hence it is identical with the Levi-Civita connection 
 with respect to $\widetilde g_{ij}$ on ${\cal M}$, and we can write
 ${\cal C}^{\downarrow}_* (\widetilde \nabla_A w_B) = \widetilde \nabla_i w_j$.   

Below we will almost exclusively use the conformally rescaled metric 
$g_{ij} = v \widetilde g_{ij} = {\cal C}^{\downarrow}_* ({V \widetilde
g_{AB}})$ on $\cal M$; in fact this extraction of $v = {\cal C}^{\downarrow}_*
(V)$ will be crucial. The covariant derivative w.r.t $g_{ij}$ is  denoted by $\nabla_i$.
   
A parametrisation of ${}^5g_{AB}$ which will be useful 
in what follows is
\begin{equation}
\label{met}
ds^2 = {}^5g_{AB}dx^A dx^B = 
V^2 \left(dx^4 + 2 A_i dx^i \right)^2 + V^{-1} g_{ij} dx^i dx^j,
\end{equation}
where $V$, $A_i$ and $g_{ij}$ depend on $x^i$ only. Such coordinates exist and will be used 
below on neighbourhoods ${\cal N}$ of compact 2-surfaces ${\cal S} \subset {\cal M}$
and of the corresponding lift ${}^5{\cal N} = {\cal C}^{\uparrow}({\cal N}) \supset {}^5{\cal S}$.  
From now onwards, all "abstract" indices materialize as coordinate indices w.r.t. (\ref{met}). 
The coordinate $x^4$ is periodic with period $Z$, i.e. $x^4$ is identified with
$x^4 + Z$.

Instead of $V$ a more common terminology is either $V = e^{2 \varphi/\sqrt{3}}$ \cite{GA}
or $V = e^{-2 \varphi/\sqrt{3}}$ \cite{HW}. Our definition of $A_i$
follows practice in relativity \cite{HW,GA,WS} but differs from practice
in gauge theory at least by a factor of 2 \cite{YC1}, if not by charge-like
constants as well. 
 
The Lagrangian density on $({}^5{\cal M}, {}^5g_{AB})$ can be decomposed 
as \cite{HW,GA}
\begin{equation}
\label{EMD}
\pounds = \frac{\sqrt{{}^5g}}{16\pi Z} {}^5R =  \frac{\sqrt{g}}{16\pi}\left[R
- \frac{3}{2} v^{-2} \nabla_i v \nabla^i v - v^3 F_{ij}F^{ij} \right], 
\end{equation} 
where  
\begin{equation}
\label{F}
F_{ij} =  {\cal C}^{\downarrow}_*({}^5 \nabla_{[A} (V^{-2} c_{B]})) = 2
\nabla_{[i}A_{j]};
\end{equation}
note that the argument of ${\cal C}^{\downarrow}_*$ is indeed an invariant object. 
 
This Lagrangian determines EMD theory on $({\cal M}, g_{ij})$. As well known, the metric variable $A_i$ in (\ref{met}), 
which is the dynamic variable in (\ref{EMD}), can be gauge 
transformed via  $A_i \rightarrow A_i + \nabla_i \Lambda$ for some function
$\Lambda$, and this corresponds to the coordinate transformation $x^4 \rightarrow x^4 + 2 \Lambda$ 
in (\ref{met}).  

We now give the action of ${\cal C}^{\uparrow}_*$ explicitly in the coordinates (\ref{met}). \\ \\
{\bf Lemma 1}. Let $y_i$ and $w^i$ be  co- and contravariant vector
fields on $\left({\cal M}, g_{ij}\right)$.
On $\left({}^5{\cal M}, g_{AB}\right)$ and in adapted coordinates
(\ref{met}), we define $ y_A = \left(y_i, 0 \right)$, $w^A = \left(w^i, 0 \right)$ and $\alpha = 2 w^iA_i$. 
Then the lifts ${\cal C}^{\uparrow}_*(y_i)$ and  ${\cal C}^{\uparrow}_*(w^i)$
are given in these coordinates by
\begin{eqnarray} 
\label{liftdn}
{\cal C}^{\uparrow}_*(y_i) & = & y_A, \\ 
\label{liftup} 
{\cal C}^{\uparrow}_*(w^i) & = & z^A := w^A - {\cal C}^{\uparrow}_*(\alpha) c^A
. 
\end{eqnarray}
{\bf Proof.} It is easy to see that the vector fields $y_A$ and $z^A$ are
invariant, viz. $ y_A c^A$, $z^Ac_A = 0$ and ${\cal L}_c y^A = 0$, ${\cal L}_c z^A = 0$, 
and they obviously project down to $y_i$ and $w^i$, i.e. ${\cal
C}^{\downarrow}_*(y_A) = y_i$ and ${\cal C}^{\downarrow}_*(z^A) = w^i$. The assertion now follows 
since ${\cal C}^{\uparrow}_*$ is an isomorphism. \\ \\
{\bf Remark.} The preceding exposition is consistent with the fibre bundle formulation of 
gauge theory \cite{YC1}. In the general setting, the fibres are not necessarily
orbits of isometries.
This entails, in particular, that there is {\it a priori} no natural "lift" of a vector field, 
but any connection on the bundle specifies a "horizontal lift". In the present
case, however, Killing transport provides a natural connection, and our lift is indeed 
"horizontal" in this sense. 
What we have done in (\ref{liftdn}) and (\ref{liftup}) now reads, in gauge theoretic language, 
to decompose the horizontal lifts of  vector fields in terms of a "direct product basis"
(\ref{met}). Applied to the basis vectors themselves, this corresponds to the 
introduction of a "gauge covariant derivative" ( cf Sect. 5 of \cite{YC1}).

\section{Horizons}

\subsection{Killing horizons}

We recall here a known result (cf. e.g. Sect. 4 of \cite{WS}) \\ \\
{\bf Theorem 1}. 
In an Einstein-Maxwell-dilaton spacetime $({\cal M},g_{ij})$ let
 ${\cal H}$ be a Killing horizon with generator $l^i$ and surface gravity
$\kappa$. That is to say, ${\cal H}$ 
is an embedded 3d null surface and $l^i$ is a Killing field defined in a neighbourhood of ${\cal
H}$ and future directed and null on ${\cal H}$ where it satisfies $l^i \nabla_i l^j = \kappa
l^j$.
Assume furthermore that ${\cal L}_l A_i = 0$ and ${\cal L}_l v = 0$.
Then the lift ${}^5{\cal H} = {\cal C}^{\uparrow}({\cal H})$ is a Killing 
horizon in the 5d vacuum spacetime $({}^5{\cal M}, {}^5g_{AB})$ with generator
$ h^A = l^A - \varphi c^A$  where $l^A= (l^i, 0)$  and $\varphi =  2 A_il^i|_{\cal H}$. 
Moreover, the surface gravity ${}^5\kappa$ defined via $h^A ~{}^5\nabla_A h^B =
{}^5 \kappa h^B$ on ${}^5{\cal H}$ agrees with $\kappa$. 
Finally, on ${}^5{\cal H}$, $h^A$ coincides with the lift 
${\cal C}^{\uparrow}_*(l^i)=  l^A - \Phi c^A$ where $\Phi = {\cal
 C}^{\uparrow}_*(\phi)$ and $\phi = 2 A_i l^i$. \\ 

We recall that $\kappa$ and $\varphi$ are constant on the horizon \cite{BC}; 
However, these constants are not  well defined 
as $l^i$ is only fixed up to a multiplicative constant, and since, moreover, 
there is the gauge freedom $A_i' = A_i + \nabla_i \Lambda$.
However, as also well known,  there is a natural way of fixing these constants 
in the asymptotically flat setting. Such a fixing is not necessary in the
present context.\\

{\bf Proof of Theorem 1.} 
We first note that $h^A$ is null on ${}^5{\cal H}$: 
\begin{equation}
\label{h2}
 h^A h_A|_{\cal H} = \left[ v^2 \left( \varphi - 2 A_i l^i \right)^2  + v^{-1} g_{ij} l^i l^j
 \right]_{\cal H} = 0.
\end{equation}
Next, $l^A$ is a Killing vector as $l^i$ leaves invariant all components of
${}^5 g_{AB}$ (cf (\ref{met})). Since $h^A$ is a constant linear combination
of two Killings, it is Killing as well. The remaining statements of the theorem 
now follow from
\begin{eqnarray} 
h^A ~{}^5\nabla_A h_B & = &- \frac{1}{2}~ ^5\nabla_B (h^A h_A) =  - \frac{1}{2} {\cal C}^{\uparrow}_*
\left(\nabla_i(v^{-1} l^jl_j)\right) = \nonumber \\
& = & {\cal C}^{\uparrow}_* \left( v^{-1} l^j \nabla_j l_i \right)=  
\kappa {\cal C}^{\uparrow}_* \left(v^{-1} l_i \right) = \kappa h_B 
\end{eqnarray}
on ${}^5{\cal H} = {\cal C}^{\uparrow}({\cal H})$, where we have used adapted coordinates 
(\ref{met}) and Eq. (\ref{h2}). $\Box$  \\ \\
{\bf Remark}. Using the formalism developed and exposed in \cite{AK,GJ,IB,MM}, it should be
possible to  generalize this Theorem to cover isolated/non-expanding/non-evolving horizons for which 
the null vectors $l^i$ and $h^A$ are only Killing on ${\cal H}$ and ${}^5{\cal H}$ respectively, 
but not necessarily in a neighbourhood. We do not go into details here.  

\subsection{The expansion} 
We now consider an orientable spacelike 2-surface ${\cal S} \subset {\cal M}$.
As all subsequent considerations will be local near ${\cal S}$, we can in
fact restrict ${\cal M}$ to a neighbourhood of ${\cal S}$. 
We introduce future directed null vectors $l^i$ and $k^i$, not necessarily Killing but orthogonal to ${\cal S}$ and scaled such that $l^i$ $k_i = -2$.
These vectors can be extended off ${\cal S}$ by considering the null geodesics with these tangents. 
This allows to define the expansion of $l^i$ (or of the corresponding null geodesics) on ${\cal S}$ via  
\begin{equation}
\label{exp}
\Theta = \left (g^{ij} + l^{(i}k^{j)} \right) \nabla_{i} l_{j}.
\end{equation} 
A standard result (cf e.g. Theorem 3.6.1 of \cite{JJ}) reads that $\Theta$ is in fact independent of the way how 
$l^i$ has been extended. 

An alternative way of writing the expansion is the following. Let ${\cal S}_{\lambda}$ 
be a family of orientable 2-surfaces in a neighbourhood of ${\cal S}$, and let $\xi^i$ be
the "lapse" of the foliation, i.e. the normal component of the corresponding 
flow vector. Moreover, let $\eta_{\lambda}$ be the volume form of ${\cal
S}_{\lambda}$ and $H^i$ the mean curvature vector \cite{JJ,PL} of ${\cal S}$.
Then (cf Sect. 1 of \cite{PL})
\begin{equation}
\label{Lxv}
{\cal L}_{\xi} \eta_{\lambda} = \xi_i H^i \eta_{\lambda}.
\end{equation}
(This formula appears more often in integrated form as "variation of area".)
 When $\xi^i$ coincides with the null direction $l^i$,
(\ref{Lxv}) reduces to 
\begin{equation}
\label{Llv}
{\cal L}_{l} \eta_{\lambda} = \Theta \eta_{\lambda}
\end{equation}
which will be used below.

\begin{figure}[h!]
\begin{psfrags}
\psfrag{M}{\Huge ${\cal M} $}
\psfrag{5M}{\Huge ${}^5{\cal M}$}
\psfrag{S}{\Huge ${\cal S} $}
\psfrag{5S}{\Huge ${}^5{\cal S} = {\cal C}^{\uparrow}({\cal S})$}
\psfrag{l}{\Huge$l^i$}
\psfrag{k}{\Huge$k^i$}
\psfrag{L}{\Huge$l^A = (l^i,0)$}
\psfrag{K}{\Huge$k^A = (k^i,0)$}
\psfrag{c}{\Huge$c^A$}
\psfrag{H}{\Huge $h^A = {\cal C}^{\uparrow}_*(l^i)$}
\psfrag{J}{\Huge $j^A = {\cal C}^{\uparrow}_*(k^i)$}
\includegraphics[angle=0,totalheight=5cm]{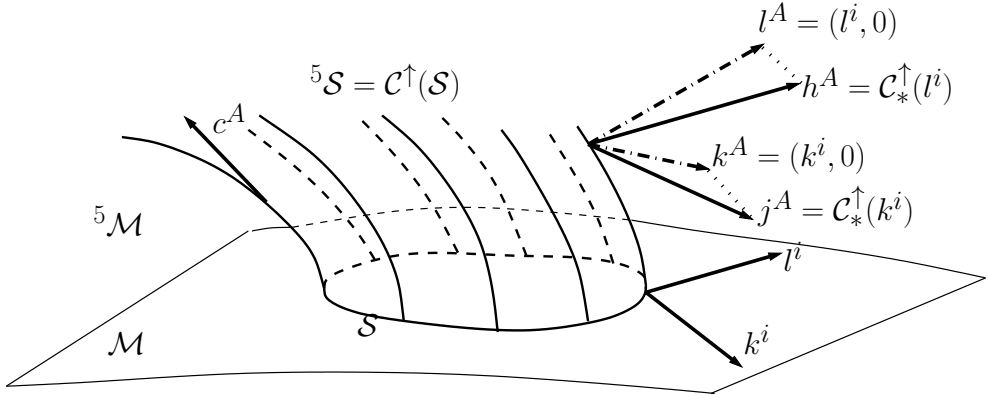}
\end{psfrags}
\caption{Lifts of null normals}
\label{lnn}
\end{figure}

For EMD theory (\ref{EMD}) we define $\psi = 2 A_i k^i$ and
we recall that $\phi = 2 A_i l^i$. These quantities will in general not be constant on ${\cal S}$.
We now lift the ${\cal S}_{\lambda}$ to 3-surfaces ${}^5{\cal S}_{\lambda} = {\cal C}^{\uparrow}({\cal
S}_{\lambda})$, define  $\Phi = {\cal C}^{\uparrow}_*(\phi)$ and $\Psi = {\cal C}^{\uparrow}_*(\psi)$, 
and lift $l^i$ and $k^i$ to invariant null vectors ${\cal C}^{\uparrow}_*(l^i) = h^A = l^A -  \Phi c^A$ 
and ${\cal C}^{\uparrow}_*(k^i) = j^A = k^A -  \Psi c^A$ on ${}^5 {\cal M}$
(cf Fig. \ref{lnn}).
These vectors are orthogonal to ${}^5{\cal S}$ since they are orthogonal to
$c^A$, and since for all tangents $m^i$ to ${\cal S}$ and  $m^A =  {\cal C}^{\uparrow}_*(m^i)$ to ${}^5{\cal S}$ we have 
$0 = {\cal C}^{\uparrow}_*(m_i l^i) = m_A h^A$  and $0 = {\cal C}^{\uparrow}_*(m_i k^i) = m_A j^A$.  
We also note that ${}^5g_{AB}h^Aj^B = - 2 V^{-1}$, and in the same way as in (\ref{exp})
and (\ref{Lxv}) we define the
expansions of $h^A$:
\begin{eqnarray}
\label{5exp1}
{}^5\Theta & = & \left ({}^5g^{AB} +  V h^{(A}j^{B)} \right) {}^5 \nabla_{A}
h_{B},\\
\label{5exp2}
{\cal L}_{h}~ {}^5\eta_{\lambda} & = & {}^5\Theta~ {}^5 \eta_{\lambda},
\end{eqnarray} 
where ${}^5 \eta_{\lambda}$ is the volume element of ${}^5 {\cal S}$.

Alternatively, we could of course start from  3d surfaces ${}^5 {\cal S}_{\lambda} \subset {}^5 {\cal M}$ 
which are invariant under an isometry.
They project to  2d surfaces ${\cal S}_{\lambda} \subset {\cal M}$, and the null normals and  expansions are
related as above.  

We now have the following result which we prove in two ways.\\ \\
{\bf Theorem 2}. Let  ${\cal S} \subset {\cal M}$ and ${}^5{\cal S} \subset {}^5{\cal M}$ 
be orientable spacelike surfaces such that ${\cal S} = {\cal C}^{\downarrow} ({}^5{\cal
S}) = {}^5{\cal S}/{\cal C}$ where 
${\cal C}$ is an isometry. Let $l^i$ and $h^A = {\cal C}^{\uparrow}_*(l^i)$ be
the respective null normals related by the Geroch-isomorphism. 
Then the corresponding null expansions defined by (\ref{exp}) or (\ref{Llv}) 
and (\ref{5exp1}) or (\ref{5exp2})
coincide in the sense that ${\cal C}^{\downarrow}_* \left({}^5\Theta(h)\right) = \Theta(l)$. \\ \\
{\bf First proof.} We manipulate (\ref{5exp1}) till it projects down to (\ref{exp}).
For ease of notation, we suppress the
 ${\cal C}^{\downarrow}_*-$ notation in both subsequent proofs. 
However, we have to take care to avoid ill-defined expressions
such as ${\cal C}^{\downarrow}_* \left( {}^5 \nabla_{A} h_{B} \right)$,
which is accomplished by appropriate algebraic decomposition of ${}^5\nabla_A h_B$. 
Next, to obtain (\ref{nabt}) we have used (\ref{cd}) while the final
step which leads to (\ref{nab}) uses the behaviour of the 4d covariant derivative under
conformal rescalings of the metric,

\begin{eqnarray}
{}^5\Theta(h) & = & \left ({}^5g^{AB} + V h^{(A}j^{B)} \right) {}^5 \nabla_{A} h_{B} = \\
& = & \left ({}^5g^{AB} + V h^{(A}j^{B)} \right) \left(P_{~A}^C +  V^{-2} c_A c^C \right)
\left(P_{~B}^D +
V^{-2} c_B c^D \right) {}^5 \nabla_{C} h_{D} =
\\ & = & \left ({}^5g^{AB} + V h^{(A}j^{B)} \right) P_{~A}^C P_{~B}^D ~{}^5 \nabla_{C} h_{D} + V^{-2} c^A c^B ~{}^5 \nabla_{A} h_{B}
= \\ & = &  \left(\widetilde g^{ij} + v l^{(i}k^{j)} \right)
\widetilde\nabla_{i}( V^{-1} l_{j}) - v^{-2} c^A h^B ~{}^5 \nabla_{A} c_{B}
= \label{nabt} \\ & = & v \left(g^{ij} +  l^{(i}k^{j)} \right) \widetilde\nabla_{i}(v^{-1} l_{j})
+ v^{-2} c^A h^B ~{}^5\nabla_{B} c_{A} = \\ 
& = & \left(g^{ij} +  l^{(i}k^{j)} \right) \widetilde\nabla_{i}l_{j} + v^{-1} l^i \nabla_i v
 = \\ & = &  
\left(g^{ij} +  l^{(i}k^{j)} \right)\nabla_{i}l_{j} -  v^{-1} l^i \nabla_i v  + v^{-1} l^i \nabla_i v =
 \Theta(l).
\label{nab} 
\end{eqnarray}
{\bf Second Proof}. Here the goal is to project  (\ref{5exp2}) into
(\ref{Llv}). 
We first note that the volume forms are related via
\begin{equation}
\label{ie}
\eta_{\lambda} = v \widetilde \eta_{\lambda} = v i_n ~{}^5  \eta_{\lambda} =
i_c ~ {}^5  \eta_{\lambda},  
\end{equation}
where $\widetilde \eta_{\lambda}$ are the 2-surface elements on ${\cal S}$ arising from the
conformally rescaled ambient metric $\widetilde g_{ij} = v^{-1} g_{ij}$, 
and $i_n$ and $i_c$ are inner products  w.r.t. the unit vector $n^A = V^{-1} c^A$ and
w.r.t  $c^A$, respectively. 
From the definitions (\ref{Llv}), (\ref{5exp2}), from (\ref{ie}) and since $h^A$ commutes with $c^A$ 
it now follows indeed that
\begin{equation}
\Theta ~\eta_{\lambda} = {\cal L}_l ~\eta_{\lambda} = {\cal L}_l \left(i_c ~{}^5\eta_{\lambda}
\right) = {\cal L}_h \left(i_c ~{}^5\eta_{\lambda}\right) = i^c {\cal L}_h {}^5 \eta_{\lambda}
={}^5 \Theta ~i^c~ {}^5 \eta_{\lambda} = {}^5\Theta ~\eta_{\lambda}.
\end{equation}
\rightline{$\Box$} \\ \\
{\bf Remark.} 
The correct conformal scaling as in (\ref{met}) is crucial both for the Kaluza-Klein decomposition 
(\ref{EMD}) as well as for the previous result. Only in the special case where $l^i$ leaves $v$ invariant, 
i.e. $l^i \nabla_i v = 0$, the conformal scaling is irrelevant. \\ 

The previous result suggests the following definition. \\ \\
{\bf Definition 1.} We define a {\bf MOTS ${\cal S} \subset {\cal M}$} to be a compact,
connected, orientable and embedded 2-surface with vanishing null expansion, i.e. $\Theta(l) = 0$. 
Analogously, a {\bf MOTS ${}^5{\cal S} \subset {}^5{\cal  M}$} is a compact,
connected, orientable and embedded 3-surface with ${}^5 \Theta(h) = 0$. \\ 

We remark that there have also been considered non-orientable MOTS \cite{GG} and non-embedded MOTS 
\cite{AMS2}. The present restrictive definition is required for some results
of Sect. 4.

\subsection{Stability and symmetry of MOTS}

 With the proper conformal scaling of the ambient metric, Theorem 2 shows that a 
MOTS  ${\cal S} \subset {\cal M}$ lifts to a surface  ${}^5{\cal S} = {\cal
 C}^{\uparrow} ({\cal S}) \subset  {}^5{\cal M}$, with vanishing expansion. 
As  ${\cal S}$ was required to be embedded, the same is true for ${}^5{\cal
 S}$. The reason is that ${}^5{\cal M}$ has locally a product structure near each point 
of ${}^5{\cal  S}$; so in particular any self intersection  ${}^5{\cal S}$
would project down to a self intersection on ${\cal S}$.
It follows that  ${}^5{\cal  S}$ is also a  MOTS. On the other hand, essentially the same 
reasoning shows that an invariant MOTS ${}^5{\cal S} \subset {}^5{\cal  M}$ projects to a MOTS  
${\cal S} = {\cal C}^{\downarrow} ({}^5 {\cal S})$. 

 In view of the significance of the property of stability as outlined in the introduction, 
we wish to show that it is preserved upon projections and lifts as well. 
Intuitively, this is clear because stability of an n-dim. MOTS 
means existence of an n+1 dim. neighbourhood whose interior and exterior
parts can be foliated by outer trapped and outer untrapped surfaces, respectively,
which means that the expansions of the leaves have appropriate signs. 
Such a foliation should then project and lift as well, 
and these manipulations should preserve the respective expansions due to Theorem 2. 
However, the details of this argument which lead to Theorem 3 involve subtleties.    
In particular, we will need two definitions of stability given in
\cite{AMS1,AMS2}. We recall below, for 2d MOTS ${\cal S} \subset {\cal M}$, 
the definition which was called "(strictly) stably outermost" in
\cite{AMS1,AMS2}, and at the risk of boring the reader we explicitly state and discuss its 
rather trivial extension to 3d MOTS ${}^5{\cal S} \subset {}^5{\cal M}$. 
The alternative definition (not repeated below) requires non-negativity (positivity)
of the principal eigenvalue of a linear elliptic "stability operator". 
These definitions are equivalent, and this equivalence will play a role in Theorem 3 below.

\begin{description}
\item[Definition 2.] 
\item[a.) Stability in 4d:] A MOTS ${\cal S} \subset {\cal M}$ is called
{\bf stable}  w.r.t. a normal direction $m^i$ iff there 
exists a variation $\delta_{\psi m}$ with $\psi \ge 0$, $\psi \not \equiv 0$ such that $\delta_{\psi m} \Theta \ge 0$.
The MOTS is strictly stable  iff it is stable with $\delta_{\psi m} \Theta \not\equiv 0$.  
\item[b.) Stability in 5d:] A MOTS ${}^5{\cal S} \subset {}^5{\cal M}$ is called stable  w.r.t. a
normal direction $n^A$ iff there exists a variation $\delta_{\Psi n}$ with $\Psi \ge 0$, $\Psi \not \equiv 0$ such that
$\delta_{\Psi n} {}^5\Theta \ge 0$.
The MOTS is strictly stable  iff it is stable with $\delta_{\Psi n} {}^5\Theta \not\equiv 0$.  
\end{description}

 {\it A priori} a MOTS  ${}^5{\cal S} \subset {}^5{\cal M}$ need not be invariant, 
nor need the direction of stability be invariant under the isometry.
However, these invariance conditions will be key in the subsequent results, and in
the next section the isometry group $U(1) \times U(1)$ will be considered.
This motivates the following definition. 

\begin{description}
\item[ Definition 3.] 
\item[a.) Stability w.r.t. an invariant direction in 4d:]
A MOTS  ${\cal S} \subset {\cal M}$ which is invariant
under an isometry  ${\cal C}$  is called {\bf stable with respect to an invariant
normal direction
$m^i$} if Definition 2 a.) applies and if ${\cal L}_c m^i = 0$.
\item[b.) Stability w.r.t. an invariant direction in 5d:]
A MOTS ${}^5{\cal S} \subset {}^5{\cal M}$ which is invariant
under an isometry group $G$  is called {\bf stable with respect to an invariant
normal direction
$\mathbf n^A$} if Definition 2 b.) applies and if ${\cal L}_g n^A = 0$ for all elements
$g^A$ of the Lie algebra ${\cal G}$ of G.
\end{description}

These definitions apply  to any normal directions $m^i$ or $n^A$. 
However, Theorem 3 below requires the {\it outgoing} conditions 
$m_i l^i > 0$ or $n_A h^A > 0$ (as it uses results of \cite{AMS2});
note that this condition still admits directions of any causal character. 
On the other hand, the area inequalities considered in Sect. 4  require in
addition that $m^i$ and $n^A$ are {\it achronal} (spacelike or null). 
All subsequent results apply if $m^i$ and $n^A$ are {\it achronal and outgoing}.
The limiting cases are $m^i = l^i$ or $n^A = h^A$  and $m^i = -k^i$ or $n^A = -j^A$. 
In the former case the variations considered above reduce to the Raychaudhuri equation and the 
stability conditions become highly restrictive, while the latter choices lead to the least restrictive conditions.    

We remark that in Definition 3 the requirement that the MOTS itself is
invariant could be relaxed once the stability condition is imposed. 
In fact there is the following relation between these two invariance properties: 
If the MOTS is  stable in a normal direction $n^A$, then any Killing field $d^A$
on ${}^5{\cal M}$ is either tangent to the MOTS
 or its normal component $d^A_{\perp}$ necessarily lies in the conical segment between the null 
generator $h^A$ of the MOTS and $n^A$  (cf. Theorem 8.1. of \cite{AMS2} and the subsequent discussion). 
Loosely speaking, a stable MOTS is automatically invariant under symmetries
of the ambient space provided the stability direction cooperates.
However, we do not exploit this fact in any of the subsequent results in order to keep
their formulation simple.   
\\ \\
{\bf Theorem 3.}  
\begin{enumerate}
\item Let ${\cal S} \subset {\cal M}$ be a MOTS which is (strictly) stable
w.r.t. an outgoing normal direction $ m^i$. Then the lifted MOTS ${\cal C}^{\uparrow}({\cal S}) \subset {}^5{\cal M}$ is (strictly) stable 
w.r.t. the outgoing normal direction $ n^A = {\cal C}^{\uparrow}_*(m^i)$.
\item Let ${}^5{\cal S} \subset {}^5{\cal M}$ be an invariant MOTS which is (strictly) stable
w.r.t. the invariant outgoing normal direction $ n^A$.
Then the projected MOTS ${\cal S} = {\cal C}^{\downarrow} ({}^5{\cal S}) \subset {\cal M}$ is 
(strictly) stable w.r.t. the normal direction $ m^i = {\cal C}^{\downarrow}_* (n^A)$.  
\end{enumerate}
~\\ \\
{\bf Proof}. 
\begin{enumerate}
\item
This is the easy part:  Let $\psi$ be the function in the definition of stability of ${\cal S} \subset {\cal M}$ 
such that $\delta_{\psi m} \Theta \ge 0$. Then the lifts $\Psi = {\cal C}^{\uparrow}_* (\psi)$
and $n^A = {\cal C}^{\uparrow}_*(m^i)$ satisfy the requirements for stability on ${}^5{\cal S} \subset {}^5{\cal M}$, 
in particular $\delta_{\Psi n} {}^5\Theta \ge 0$. More explicitly, in
terms of a coordinate chart $x^i$ near ${\cal S}$, and an adapted coordinate $z$ such that $\psi m^i \partial/\partial x^i =
\partial/\partial z$,  we have  $\delta_{\psi m} \Theta = \partial/\partial
z~ \Theta\ge 0$. Lifting all coordinates to a neighbourhood of ${}^5 {\cal
S}$ we obtain the result.     

\item
In spite of the invariance assumptions there is still a subtlety here: The function $\Psi$ which defines the variation 
$\delta_{\Psi n} {}^5\Theta \ge 0$ in Definition 3.b.) of stability of ${}^5{\cal S}$ need not be invariant under ${\cal C}$, 
in which case it does not project to ${\cal S} \subset {\cal M}$. Here we need the elliptic machinery developed in \cite{AMS1,AMS2}. 
We rewrite the variation $\delta_{\Psi n} {}^5\Theta$ in terms of a linear elliptic operator ${}^5L_n$ acting on $\Psi$ such that
$\delta_{\Psi n} {}^5\Theta = {}^5L_n \Psi$.  Any linear elliptic operator has a real, positive  "principal" eigenfunction
$\Phi$ corresponding to the "principal" eigenvalue $\lambda$ (the eigenvalue with lowest real part), viz. ${}^5 L_n \Phi = \lambda \Phi$. 
For ${}^5L_n$ introduced above, the definitions of stability and strict stability are equivalent 
to $\lambda \ge 0$ and $\lambda > 0$, respectively. This means in particular that for a stable MOTS, 
the principal eigenfunction  $\Phi$ defines a prefered class of variations such that 
$\delta_{\Phi n} {}^5 \Theta = {}^5L_n \Phi = \lambda \Phi \ge 0$.  (or $ > 0$ in the strictly stable case). 
But by virtue of the invariance of the stability direction $n^A$, the stability operator commutes with the isometry, viz. 
${\cal L}_c {}^5L_n = {}^5 L_n {\cal L}_c$. Now Theorem 8.2. of \cite{AMS2} implies that the principal 
eigenfunction is invariant under the isometry, i.e. ${\cal L}_c \Phi = 0$. Hence $\Phi$ can be projected to ${\cal M}$ 
and $\phi = {\cal C}^{\downarrow}_* \Phi$ defines the variations on ${\cal S}$ which imply stability, 
in particular $\delta_{\phi m} \Theta \ge 0$. 
\end{enumerate}
\rightline{$\Box$}

\section{Area inequalities}
Area inequalities bound the area $A$ of stable MOTS in terms of quantities defined on the MOTS,
namely charges and, in the axially symmetric case, angular momenta  (\cite{SD,JLJ2}).
Here we first briefly recall the area inequalities for
Einstein-Maxwell  and Einstein-Maxwell-dilaton in 4d, and in a 5d vacuum. 

For axially symmetric, stable 2d MOTS in Einstein-Maxwell theory, 
Gabach Clement et al. \cite{GCJ,GJR} proved that 
\begin{equation}
\label{GJR}
A \ge 4\pi \sqrt{ 4 J^2 + (Q^2 + P^2)^2},
\end{equation} 
where $J$ is the angular momentum and $Q$ and $P$ are the electric and
magnetic charges. Equality holds iff the near horizon geometry is an extreme Kerr-Newman one. 

In EMD theory, for couplings which include Einstein-Maxwell (i.e. no dilaton)
as well as the Lagrangian (\ref{EMD}),  
Yazadjiev has shown (Theorems 1 and 2 of \cite{SY}) that for 2d MOTS
which are stable w.r.t. an achronal, outgoing normal direction,
\begin{equation}
\label{SY}
A \ge 8\pi \sqrt{ \left| J^2 - Q^2 P^2 \right|}.
\end{equation}
Moreover, for the KK coupling equality holds iff the (near-) horizon geometry
\cite{KL,HI2} is extreme and stationary.

On the other hand, in a 5d vacuum spacetime with isometry group
$U(1)^2 = U(1) \times U(1)$  Hollands has
obtained an area inequality for invariant 3d MOTS ${}^5{\cal S}$ which are
stable  w.r.t. an achronal, outgoing, invariant normal direction
(Theorem 1 of \cite{SH}). It takes the form 
\begin{equation}
\label{SH}
{}^5 A\ge 8\pi \sqrt {\left |{}^5J_+~ {}^5J_- \right|}, 
\end{equation}     
where ${}^5J_+$ and ${}^5 J_-$ are angular momenta associated with the isometries.
Again equality holds precisely for the stationary, extreme (near-) horizon
geometries.

We now compare these inequalities. Regarding the relation between (\ref{GJR}) and (\ref{SY}) the situation is clear: 
As already noticed in \cite{SY}, the former is the stronger inequality but it only applies
in the Einstein-Maxwell case. On the other hand, (\ref{SY}) and (\ref{SH}) should have a common range of applicability:
As we have discused at length, a 5d vacuum with a spacelike isometry is equivalent to EMD theory in
4d, and the isometry preserves stable MOTS in the sense of Theorems 2 and 3.

However, regarding invariance, stability and topology the setup of \cite{SY} and \cite{SH} is 
different and in fact also different from our favourite setup as described in Sect. 3.3.
We have to clarify these issues before investigating the relation between the parameters.

\begin{description}

\item[Invariance of the ambient space.]  
Eqs. (\ref{SY}) and (\ref{SH}) hold, respectively, for MOTS which are invariant
under the $U(1)$ and $U(1)^2$ symmmetry. However, invariance of the full 
ambient geometry is assumed in \cite{SH} but not in \cite{GCJ,GJR,SY}.
To keep the following discussion simple we always assume below symmetry of the
full ambient geometry.  

\item [Invariance of the MOTS.]    
In \cite{SH} where $U(1)^2$ invariance of the ambient geometry is assumed,
it is stated that the considered MOTS ${}^5{\cal S}$ is then automatically invariant under this symmetry. 
This is due to the very {\it construction} of that MOTS via an appropriate  cross section of an 
invariant light cone.
Here we assume, in contrast, that a MOTS ${}^5{\cal S}$ is {\it given}, and accordingly we
 {\it require} henceforth its invariance under the ambient symmetry group. 
(See, however, the remark after Definition 3 regarding a possible relaxation of this
 requirement for {\it stable} MOTS).      

\item[Invariance of the stability direction.] 
The  proofs \cite{GCJ}--\cite{SH} of the area inequalities 
(\ref{GJR}) -- (\ref{SH}) agree more or less  
regarding the assumptions of compatibility of the stability direction
with the symmetries. All these assumptions are formulated in terms of
{\it scalar functions} with respect to suitably scaled null bases.
However, in view of the discussion
of the previous section, we prefer to set out from our purely geometric 
Definition 3  instead. Then the proof of the second part of Theorem 3
shows that there exist invariant lapse functions $\phi$ and $\Phi$.
This implies in particular stability in the sense used in the proofs 
of the area inequalities.
   
\item[Topology of the MOTS.] The MOTS ${}^5{\cal S}$ are 3-dimensional, connected, 
orientable, compact Riemannian manifolds with a two-dimensional isometry group containing
$U(1)^2$. It has been shown in Theorem 2 of  \cite{HY} that such surfaces
are topologically one of $\mathbb{S}^3$, $\mathbb{S}^2 \times \mathbb{T}$, 
the lens spaces $L(p,q)$,  or $\mathbb{T}^3$ (where $p,q \in \mathbb{Z}$ with g.c.d.
$(p,q) = 1$, and  $\mathbb{T}$ denotes a closed curve). 
The $\mathbb{T}^3$ topology is in fact incompatible with the stability 
condition \cite{GG} and will therefore not be considered further.  
On the other hand, (\ref{SY}) holds for MOTS of topology $\mathbb{S}^2$.
Therefore, a necessary prerequisite for obtaining (\ref{SY}) from (\ref{SH}) upon dimensional 
reduction is that we set out from a MOTS ${}^5{\cal S}$ which is a $\mathbb{T}$ bundle over $\mathbb{S}^2$. 
Such bundles have been classified \cite{NS} and they clearly include 
$\mathbb{S}^3$ and $\mathbb{S}^2 \times \mathbb{T}$. As to the Lens spaces $L(p,q)$ for $p\ge 2$,
 $q\ge 2$, they involve discrete identifications in two directions. While one of
 these directions can be
 aligned with the ${\mathbb T}$- fibre, the other one would lead to "orbifold-"
 identifications on $\mathbb{S}^2$ which we do not consider here. 
For the Lens spaces $L(p,1)$, however, only the fibre identification remains,
 whence this case is admitted below as well.  For the $\mathbb{S}^2$ bundles, Theorem 2 of \cite{HY} also
 shows that there is a "Kaluza-Klein-" subgroup $U(1) \subset U(1)^2$ which acts 
freely on ${}^5{\cal S}$ and is aligned with the $\mathbb{T}$- fibres. 

Note that  (\ref{SH}) holds for all lens spaces $L(p,q)$, not necessarily sphere
bundles. Hence this inequality is more general than (\ref{SY}) in this sense. 

\end{description}

We proceed with two Theorems. Theorem 4 just reviews key elements of the lift and projection procedures 
which are in particular needed to relate the requirements used in the proofs of (\ref{SY}) and
(\ref{SH}). Its proof follows readily from Theorems 2 and 3 and the preceding remarks.
In Theorem 5 we then define and analyze the parameters occuring in 
the area inequalities. \\ \\
{\bf Theorem 4}.
\begin{enumerate}
\item
Let $({\cal M}, g_{ij}, A_i, v)$ be an axially symmetric EMD spacetime containing a 
2-dim MOTS ${\cal S}$ with "axial" isometry ${\cal A}$ (two fixed points) 
and corresponding Killing vector $\xi^i$.  
Then lifting to a 5-dim vacuum spacetime  $({}^5{\cal M}, {}^5g_{AB})$ with periodic isometry 
${\cal C}$ and corresponding Killing vector $c^A$ gives an ${\cal A} \times {\cal C}$ invariant MOTS ${}^5{\cal S}$
with additional symmetry $\Xi^A = {\cal C}^{\uparrow}_*(\xi^i)$. 
If ${\cal S}$ is stable w.r.t. an outgoing ${\cal A}$-invariant normal direction $m^i$ in the sense of Definition 3, then
${}^5{\cal S}$ is stable w.r.t. to the ${\cal A} \times {\cal C}$-invariant
direction ${\cal C}^{\uparrow}_*(m^i)$.
  
\item Conversely, let $({}^5{\cal M}, {}^5g_{AB})$ be a 5-dim vacuum spacetime 
with a $U(1)^2$ isometry. Let ${}^5{\cal S} \subset {}^5{\cal M}$
be a MOTS which is topologically a $\mathbb{T}$ bundle over $\mathbb{S}^2$ 
(hence either $\mathbb{S}^3$, $\mathbb{S}^2 \times \mathbb{T}$ or $L(p,1)$)
and which is invariant under the ambient isometries.
Then there exist two $U(1)$ subgroups ${\cal C}$ and ${\cal A}$,
 the former  acting freely on ${}^5{\cal M}$, 
with corresponding Killing fields $c^A$ and $\Xi^A$.  
Dimensional reduction then yields an axially symmetric EMD spacetime $({\cal M}, g_{ij}, A_i, v)$ 
containing a 2-dim, axially symmetric MOTS ${\cal S}$ with axial Killing vector $\xi^i = {\cal C}^{\downarrow}_*(\Xi^A)$. 
Moreover, if ${}^5{\cal S}$ is stable in an outgoing direction $m^A$ which is invariant under 
$U(1) \times U(1)$, ${\cal S}$ is stable in a direction ${\cal
C}^{\downarrow}_*(m^A)$ which is invariant under the axial isometry.  
\end{enumerate} 
{\bf Theorem 5.}
 Let  the spacelike surface ${}^5{\cal S} \subset {}^5{\cal M}$ 
be a bundle of topology $\mathbb{S}^3$,  $L(p,1)$ or  $\mathbb{S}^2 \times
 \mathbb{T}$,  such that the base
 ${\cal S} = {\cal C}^{\downarrow}({}^5{\cal S})= {}^5{\cal S}/{\cal C}$ has topology $\mathbb{S}^2$ and the fibres ${\cal C}$ are isometries. 
Then the areas of ${}^5{\cal S}$ and ${\cal S}$ ( w.r.t. the measures 
${}^5\eta$ and $\eta$) are related by ${}^5A = Z A$  
(where $Z$ is the periodicity of $x^4$), 
and charges $Q$ and $P$ are defined and related via (\ref{Q}) and (\ref{P})
\begin{eqnarray}
\label{Q}
Q & = & \frac{1}{4\pi} \int_{\cal S} v^3 F_{ij} dS^{ij} = 
\frac{1}{4\pi Z} \int_{{}^5{\cal S}}
{}^5\nabla_{A}c_{B} dS^{AB}, \\
\label{P}
P & = & \frac{1}{4\pi} \int_{\cal S} * F_{ij} dS^{ij} = 
\frac{1}{8\pi Z} \int_{{}^5{\cal S}} \epsilon^{ABC}_{~~~~~DE} V^{-2} c_A {}^5\nabla_{B}
(V^{-2} c_{C}) dS^{DE}. 
\end{eqnarray}
Moreover, $8\pi P = Z p$ where $p = 0$ for topology $\mathbb{S}^2 \times
\mathbb{T}$, $p = 1$
for $\mathbb{S}^3$ and $p$ agrees with the lens space parameter otherwise. 

Furthermore, if  ${\cal S} \subset {\cal M}$ and ${}^5{\cal S} \subset {}^5{\cal M}$   
are axially symmetric, their angular momenta are 
\begin{eqnarray}
\label{J}
J & = & \frac{1}{8\pi} \int_{\cal S} \nabla_{i} \xi_{j} dS^{ij} = 
\frac{1}{8\pi Z} \int_{{}^5{\cal S}} \nabla_{A} \Xi_{B} dS^{AB}.
\end{eqnarray}
{\bf Proof}. 
 The area $A$ of any invariant 2-surface ${\cal S}$ 
is related to the area ${}^5A$ of its lift ${}^5{\cal S} = {\cal C}^{\uparrow}_*({\cal S})$ via  
\begin{equation}
{}^5A = \int_{{}^5{\cal S}} {}^5 \eta =  \int_{{}^5{\cal S}} \widetilde\eta \wedge n  =   
\int_0^Z \int_{\cal S}  \eta \wedge dx^4 = Z ~ A,  
\end{equation}
where $\widetilde \eta$ refers to the metric $\widetilde g_{ij}$, and $n = V dx^4$ is the 1-form dual to the unit vector 
$n^A = V^{-1} c^A$.

To show the equivalence of the two representations in (\ref{Q}), (\ref{P}) and (\ref{J}) we first note that the surface 
element on ${}^5{\cal S}$ reads $dS^{AB} = V  h^{[A} j^{B]} ~{}^5\eta$ in terms of the lifts $h^A = {\cal C}^{\uparrow}_*(l^i)$ and 
$j^A = {\cal C}^{\uparrow}_*(k^i)$,  as the latter are not normalized but scale as $h^A j_A =  -2V^{-1}$,  cf Sect. 3.2.

Recalling now (\ref{F}) we find that the integrands are related as follows  
\begin{eqnarray}
{\cal C}^{\downarrow}_*\left[\left({}^5\nabla_{A}c_{B} \right) h^A j^B \right]
& = & {\cal C}^{\downarrow}_*\left[ V^2
 ~{}^5\nabla_{[A}\left( V^{-2} c_{B]}\right)\right] 
{\cal C}^{\downarrow}_*\left(  h^{A} j^{B}  \right)  = \nonumber \\
& = & 2 v^2 \left(\nabla_{[i} A_{j]}\right)  l^i k^j  =  v^2 F_{ij} l^i k^j,
\\  
{\cal C}^{\downarrow}_*\left[ \epsilon^{ABC}_{~~~~~DE} V^{-2} c_A ~{}^5\nabla_{B}
(V^{-2} c_{C}) h^D j^E \right] & = & v^{-1} \epsilon_{ij}^{~~kl} F_{kl} k^i
l^j,
\\
{\cal C}^{\downarrow}_* \left[\left( \nabla_{A} \Xi_{B} \right) h^A j^B \right] & = & v^{-1} \nabla_{i} \xi_{j} k^i l^j. 
\end{eqnarray}
Finally, to show the relation $8\pi P = Z p$ we recall a calculation
in gauge theory which appears frequently in spherically symmetric 
settings  (cf. e.g.  Example 10.1 and Sect. 10.5.2 of \cite{MN}) but which we perform here in general. 
We consider domains ${\cal U}$ and ${\cal D}$ such that  ${\cal U} \cup {\cal D} = {\cal S}$ and a smooth closed path ${\cal P} \subset  {\cal U} \cap {\cal D}$ 
with unit tangent $t^i$ and parameter length $2\pi$.  On either domain we have connection 1-forms $A^U_i$ and $A^D_i$ 
related by a gauge transformation $\Lambda :  {\cal U} \cap {\cal D}
\rightarrow U(1)$ with  $A^U_i - A^D_i = \nabla_i
\Lambda$. Upon integration, we obtain  
\begin{eqnarray}
8 \pi P & = & 2 \int_{\cal S} * F_{ij} dS^{ij} =  
2 \int_{\cal S} \epsilon_{ijkl} \nabla^{i} A^{j} dS^{kl} = \\
 & = & 2 \int_{\cal P} A^U_i t^i dS - 
 2 \int_{\cal P} A^D_i t^i dS  = 
2 \int_{\cal P} \nabla_i \Lambda t^i dS  = 2 \left[ \Lambda (2\pi) -
 \Lambda(0) \right]. 
\end{eqnarray}
Thus, when $P \neq 0$, $\Lambda$ is multi-valued on ${\cal U} \cap {\cal D}$.
We now recall  from Sect. 2 that the gauge transformation $A_i \rightarrow A_i + \nabla_i \Lambda$ 
corresponds to the motion $x^4 \rightarrow x^4 + 2 \Lambda$ in the $U(1)$
fibre. Definiteness of the connection requires that $\Lambda({\cal P})$ defines a
closed orbit on the torus ${\cal P} \times U(1)$, so that moving once around ${\cal P}$ corresponds to $p$ loops
around the fibre. In other words, there is a homotopy from 
${\cal P}$ to  $U(1)$  with winding number $p$. With a fibre of length $Z$, this gives 
$8 \pi P =  2 \left[ \Lambda (2\pi) - \Lambda(0) \right]=  Z p $ as claimed.  
We remark that the monopole $P$ also agrees with the first Chern number of the bundle
(cf. Example 11.2 of \cite{MN}).\\
\rightline{$\Box$} 

{\bf Remark}. The second representation (\ref{P}) of the magnetic monopole
provides an interesting connection with the Hopf invariant \cite{MN,BT} 
of the bundle. To see this recall that
\begin{equation}
\label{g2}
\frac{1}{8\pi P}F_{ij} = \frac{1}{4\pi P} \nabla_{[i} A_{j]}
\end{equation}
is a generator of the second cohomology group $H^2({\cal S})$ which is
non-trivial for $\mathbb{S}^2$. Hence (\ref{g2}) holds in general only
locally. On the other hand, if ${}^5{\cal S}$ has topology $\mathbb{S}^3$ or lens
space topology $L(p,1)$, the lift is exact globally, viz. 
\begin{equation}
{\cal C}^{\uparrow}_* \left[ \frac{1}{8\pi P} F_{kl} \right] =
 \left[{}^5\nabla_{[A}\left(\frac{ c_{B]}}{8\pi P V^2}\right)\right]. 
\end{equation}

In either case, the isometry ${\cal C}$ provides a map 
${\cal C}^{\downarrow}: \mathbb{S}^3 \rightarrow \mathbb{S}^2$
with Hopf invariant  
\begin{equation}
\label{H}
H({\cal C}) =  \int_{{}^5\widehat{\cal S}} \epsilon^{ABC}_{~~~~~DE} \frac{c_A}{8 \pi P V^2}  {}^5\nabla_{B}
(\frac{c_C}{8\pi P V^2}) dS^{DE}, 
\end{equation}
where  ${}^5\widehat{\cal S}$ coincides with ${\cal S}$ for $\mathbb{S}^3$
topology but is the p-fold $\mathbb{S}^3$-cover in the case of the lens space
$L(p,1)$.

We now combine this with (\ref{P}), take into account that the integral over
$ {}^5\widehat{\cal S}$ is $p$ times the
integral over ${}^5{\cal S}$ and use $8\pi P = Z p$. It follows that 
$H({\cal C}) = 1$; in particular the Hopf invariant (which is an integer for
any smooth map $\mathbb{S}^3 \rightarrow \mathbb{S}^2$) 
does not reflect the winding number $p$.
\rightline{$\Box$}

We now define Killing vectors
\begin{eqnarray}
  \Upsilon^A_{\pm}  = \pm 2 P c^A + \Xi^A =  \pm \frac{Z p}{4\pi} c^A + \Xi^A 
\end{eqnarray}
and corresponding Komar integrals
\begin{equation}
{}^{5}J_{\pm} = \frac{1}{8\pi } \int_{{}^5{\cal S}} {}^5\nabla_{A} \Upsilon_{B}^{\pm}
dS^{AB}, \\
\end{equation}
which from the definitions (\ref{Q})--(\ref{J}) gives the further relation ${}^5J_{\pm} = Z (J \pm PQ)$. 

To complete the identification of Theorems 1 of \cite{SY} and \cite{SH}
containing (\ref{SY}) and (\ref{SH}) respectively, requires two trivial inputs. 
Firstly, in \cite{SH} all Killing vectors have period
$2\pi$, while our isometry ${\cal C}$ has a (dimensional) length $Z$. 
This can of course be fixed by adjusting units.
Secondly, the integer $p$ defined in Theorem 5 matches the synonymous integer 
defined above Theorem 1 of \cite{SH}, while we have to set the integer $q$ of 
\cite{SH} equal to $1$ for the topological reason discussed above Theorem 4.

As already mentioned the extreme and stationary geometries in which 
(\ref{SY}) and ({\ref{SH}) are saturated are included in the 
"near horizon geometries" analysed in \cite{KL,HI2}. 
Among them, there is a large class of solutions which are known globally
\cite{HW}:
For the trivial bundle $p = 0$ (topology $\mathbb{S}^2 \times \mathbb{T}$), these include the 
boosted extremal Kerr and the black ring topologies, while for $p=1$ (topology 
$\mathbb{S}^3$) the geometries are either the extremal Myers-Perry or the 
extremal rotating Kaluza-Klein ones. For these stationary solutions, the
relation between the 4d and 5d parametes has been obtained and 
discussed in \cite{EM} (cf Appendix A2 in particular; again units need to be
adjusted).    

In this context we finally note that, from Theorem 1, the property of extremality ($\kappa = 0$) 
of a stationary Killing horizon is preserved upon lift and projection, 
irrespective of any additional symmetries. This behaviour is of course reflected 
in the axially symmetric examples mentioned above. Moreover, it should extend to
isolated/non-expaning/non-evolving horizons by virtue of the results of 
\cite{MM,JLJ3}.
\\ \\
{\large\bf Acknowledgements.}
We are grateful to Piotr Bizo\'n, Andreas \v Cap, Piotr Chru\'sciel, Jose-Luis Jaramillo, Marc Mars and Helmuth Urbantke 
for helpful discussions and correspondence. We also (or rather: in particular) thank the referees for
 important comments which led to substantial improvements. 
The research of W.S. was funded by the Austrian Science Fund (FWF): P 23337-N16


\begin{thebibliography}{00}
\bibitem{OW} Overduin J M and Wesson P S 1997 {\it Phys. Rep.} {\bf 283} 303  
\bibitem{YC1} Cho Y M 1975 {\it J. Math. Phys.} {\bf 16} 2029
\bibitem{ER} Emparan R and Reall H S 2008 {\it Living Rev. Rel.} {\bf 11:6}
\bibitem{HW} Horowitz G T and Wiseman T {\it in} 2012 Horowitz G T (ed.) {\it  Black Holes in Higher
Dimensions}~(Cambridge: Cambridge University Press) 
\bibitem{GA} Gibbons G 1982 {\it Nucl. Phys. B} {\bf 207} 337
\bibitem{WS} Simon W  1985 {\it Gen. Rel. Grav.} {\bf 17} 439 
\bibitem{DR} Rasheed D 1995 {\it Nucl. Phys. B}~{\bf 454} 379
\bibitem{FL} Larsen F 2000 {\it Nucl. Phys. B}~{\bf 575} 211
\bibitem{HI1} Hollands S and Ishibashi A 2012 {\it Class. Quantum Grav.}~{\bf 29} 163001
\bibitem{JLJ1} Jaramillo J L  2011 {\it Int. J. of Modern Physics D}~{\bf 20} 2169
\bibitem{AMS1} Andersson L, Mars M and Simon W 2005 {\it Phys. Rev. Lett.} {\bf 95} 111102
\bibitem{AMS2} Andersson L, Mars M and Simon W 2008 {\it Adv. Theor. Math. Phys} {\bf 12} 853
\bibitem{AM} Andersson L and Metzger J 2009 {\it Commun. Math. Phys}~{\bf 290} 941 
\bibitem{ME} Eichmair M 2009 {\it J. Diff. Geom}~{\bf 83} 551 
\bibitem{AME} Andersson L, Eichmair M and Metzger J {\it in} 2009 Agranovsky
M et al. (eds) {\it Proceedings of the Fourth International
Conference on Complex Analysis and Dynamical Systems (Contemporary Mathematics 553)}  
(Providence: American Mathematical Society and Ramat-Gan: Bar-Ilan University)
\bibitem{HE} Hawking S W and Ellis G F R 1973 {\it The large scale structure
of space-time}~(Cambridge: Cambridge University Press).
\bibitem{RN} Newman R P A C 1987 {\it Class. Quantum Grav.}~ {\bf 4} 277
\bibitem{GG} Galloway G {\it in} 2012 Horowitz G T (ed.) {\it  Black Holes in Higher
Dimensions}~(Cambridge: Cambridge University Press) 
\bibitem{AMMS} Andersson L, Mars M, Metzger J and Simon W 2009 {\it Class. Quantum
Grav.}~ {\bf 26} 085018
\bibitem{IC} Costa e Silva I P 2012 {\it Class. Quantum Grav.}~{\bf 29} 235008 
\bibitem{JRD} Jaramillo J L, Reiris M, Dain S 2011 {\it Phys Rev. D}~ {\bf 84} 121503(R) 
\bibitem{SD} Dain S 2012 {\it Class. Quantum Grav.}~{\bf 29} 073001
\bibitem{JLJ2} Jaramillo J L 2013 {\it Springer Proceedings in Mathematics \& Statistics} {\bf 26} 139 
\bibitem{GCJ}  Gabach Clement M E and Jaramillo J L 2012 {\it Phys. Rev. D 86} 064021
\bibitem{GJR} Gabach Clement M E, Jaramillo J L and Reiris M 2013 {\it Class. Quantum Grav.} (to be published)
\bibitem{SY}  Yazadjiev S 2013 {\it Phys. Rev. D}~{\bf 87} 024016  
\bibitem{SH} Hollands S 2012 {\it Class. Quantum Grav.}~{\bf 29} 065006
\bibitem{RG}  Geroch R 1971 {\it J. Math. Phys.}~{\bf 12}, 918
\bibitem{WP} Pauli W {\it in} 1955 Jordan P {\it Schwerkraft und Weltall}~(Braunschweig: F. Vieweg und Sohn)
\bibitem{YC2} Cho Y M  1992 {\it Phys. Rev. Lett.}~{\bf 68} 3133 
\bibitem{BC} Carter B {\it in} 1973 DeWitt C \& DeWitt B S (eds.) {\it Black hole
equilibrium states} (New York: Gordon \& Breach)
\bibitem{JJ} Jost J 1998 {\it Riemanian Geometry and Geometric Analysis}~
(Berlin, Heidelberg, New York: Springer) 
\bibitem{PL} Li P 2012 {\it Geometric analysis}~(Cambridge: Cambridge University Press). 
\bibitem{AK} Ashtekar A and Krishnan B 2004 {\it Living Rev. Rel.}~{\bf 7:10}
\bibitem{GJ} Gourgoulhon E and Jaramillo J L 2006 {\it Phys.Rept.}~{\bf 423} 159
\bibitem{IB} Booth I  2013 {\it Phys.Rev. D}~{\bf 87} 024008  
\bibitem{MM} Mars M 2012 {\it Class. Quantum Grav.}~{\bf 29} 145019
\bibitem{EM} Emparan R and Maccarrone A 2007 {\it Phys Rev. D}~{\bf 75} 084006
\bibitem{KL} Kunduri H K and Lucietti J 2009 {\it J. Math. Phys.}~{\bf 50} 082502 
\bibitem{HI2} Hollands S and Ishibashi A 2010 {\it Ann. Henri Poincar\'e}~{\bf 10} 1537
\bibitem{HY} Hollands S and Yazadjiev S 2011 {\it Commun. Math. Phys}~{\bf 302} 631
\bibitem{NS} Steenrod N 1951 {\it The Topology of Fibre Bundles}~(Princeton:
Princeton University Press)  
\bibitem{MN} Nakahara M 2003 {\it Geometry, Topology and Physics}~(Boca
Raton: Taylor \& Francis Group)
\bibitem{BT} Bott R and Tu L W 1982 {\it Differential Forms on Algebraic Topology}
~(Berlin, Heidelberg, New York: Springer) 
\bibitem{JLJ3} Jaramillo J L 2012 {\it Class. Quantum Grav.}~{\bf 29} 177001
\end{thebibliography}
\end{document}